\documentclass[]{amsart}
\usepackage{psfig}
\begin{document}

\title{Mass Transfer Mechanism in Real Crystals by Pulsing Laser
Irradiation}

\author{A.E.Pogorelov (a), K.P.Ryaboshapka (a)}
\address{(a) G.V.Kurdyumov Institute for Metal Physics of NAS of
Ukraine, 36 Vernadskii blrd., 03142 Kiev, Ukraine}

\author{A.F.Zhuravlyov (b)}
\address{(b) Institute for Magnetism of the NAS of Ukraine,
36-b Vernadskii blrd.,03142 Kiev, Ukraine} \email{(B)
zhur@imag.kiev.ua}


\date{\today}

\begin{abstract}
The dynamic processes in the surface layers of metals subjected
activity of a pulsing laser irradiation, which destroyed not the
crystalline structure in details surveyed. The procedure of
calculation of a dislocation density generated in bulk of metal
during the relaxation processes and at repeated pulse laser
action is presented. The results of evaluations coincide with
high accuracy with transmission electron microscopy dates. The
dislocation-interstitial mechanism of laser-stimulated mass
transfer in real crystals is presented on the basis of the ideas
of the interaction of structure defects in dynamically deforming
medium. The good compliance of theoretical and experimental
results approves a defining role of the presented mechanism of
mass transfer at pulse laser action on metals. The possible
implementation this dislocation-interstitial mechanism of mass
transfer in metals to other cases of pulsing influences is
justified
\end{abstract}

\maketitle

\section{Introduction}

It was appeared experimentally in the works \cite1-\cite3 that the
Q-switched laser pulses with power density $q_i=10^6\div 10^7
W/cm^2,$ pulse duration of $\tau=10\div 50 ns$ stimulate anomalous
deep atom penetration from irradiated metal surface to the depth
value of which is considerably more than thermal influence depth.
To interpret this phenomenon of high atomic mobility, a mechanism
of interstitial atom migration under the conditions of high-rate
strain was presented in \cite1. This mechanism cannot interpret
observed experimentally long-range mass transfer, because the
interstitial atoms are short-live defects, which quickly
eliminated on sinks. To escape this problem the authors of the
work \cite4 supposed interstitial atom dragging by the
dislocations, which were moving by the thermal stresses. In this
case lifetime of the complex {\it "dislocation+interstitial atom"}
was sufficient for right estimations of the penetration depth.
Thermal stresses were calculated using the exact solution \cite5
of the thermal conductivity problem in the laser-influenced metal.
The calculations show a possibility of generation of the
dislocations during a relaxation of the deformations in the
metal. The deformation processes during the laser actions were
confirmed by the presence of topographical curving structure and
formation of polygonal grids in crystalline $Mo$ \cite6, and
cellular dislocation structures in $Armco-Fe$ \cite7. Dislocation
structure analysis, which was performed with the data presented
in the work \cite8, has evidenced the possibility of generating
large amount of point defects of interstitial type under these
conditions. Model calculations, which were performed in the work
\cite4, show the dynamical state of the laser-influenced metal.
This state is stipulated by the moving of the gradients of thermal
rapid-changed stresses from the skin layer into the bulk of the
metal. These stresses cause the compressing waves, even under
special conditions shock wave forming, and penetration in
considerable depth of laser-influenced metals. The formation of
polygonal grids in $Mo$ \cite6, and cellular dislocation
structures in $Armco-Fe$ \cite7 demonstrate the interaction
between dynamic stresses and dislocations, which are transferred
in dynamically deformed lattice.

In the present work we discuss the problem of the creation,
annihilation, and lifetime of the {\it "dislocation+interstitial
atom"} complex in the frame of the dislocation-interstitial
mechanism in the metals, which are influenced by pulsing laser
irradiation. The results permit correct interpretation of
anomalous deep atom penetration from the laser-affected surface
to the bulk of the metal on distances of tens of micrometers.

\section{Basic relations}

Pulsing laser heating of the near-to-surface layers of metals
stipulates the rapid thermal expansion of these layers. Heat
velocity is connected with the velocity of the temperature change
with the time $t:$
$$
\partial_t T\approx 10^6\div 10^{10}  K/s$$
which, in turn, depends on the duration of the laser pulse. It
was shown in the work \cite9, that in the case of duration of
laser pulse $\tau\approx 10^{-10} s$ the stress fields in the
influenced metal one can describe to solve the dynamic
thermal-elasticity equations \cite{10} in the quasi-stationary
approach \cite{11}.

The temperature $T(r,\phi ,z,t)$ in bulk of the metal ($z>0$),
which is irradiated by the laser beam with the width $\beta,$
intensity $I_0 ,$ duration $\tau ,$ and skin layer $\kappa,$ is
described by the equation \cite{5}:

\begin{eqnarray}\label{D33.7} \nonumber
\partial_t T-a\Delta T=I_0 e^{-\beta r^2 -\kappa z} \left\{
\begin{array}{cc}
1,&{\rm if}\; t\leq \tau , \\ 0,&{\rm if}\; t>\tau ,
\end{array} \right.
\end{eqnarray}
with boundary conditions:
$$
(\partial _z T)_{z=0}=hT(r,0,t).$$ Here  $a$ is thermal
conductivity coefficient; $I_0$ is defined as
$$\frac{q_0}{k}a\eta ,$$ where $q_0$ is the power of irradiation,
$\eta$ is the absorption coefficient, $k$ is thermal capacity
coefficient; $h$ describes the heat flow absorbing; $\Delta$ is
the Laplasian. We omit below coordinate $\phi$ due to the
cylindrical symmetry of the considered problem.

Exact solution of this problem is:

\begin{eqnarray}\label{TF}
T(r,z,t)&=&\frac{I_0}{2\beta}\int\limits_0^{\infty}\xi J_0 (\xi
r)e^{-\frac{\xi^2}{4\beta}}\lbrace F(\xi ,t)- \nonumber \\ &-&
\left\{\begin{array}{cc} 0,&{\rm if}\quad t\leq\tau \\ F(\xi
,t-\tau ),&{\rm if}\quad t>\tau
\end{array}\right\} d\xi .
\end{eqnarray}
where
\begin{eqnarray}\label{F}
F(\xi ,t)&=&\frac{1}{a(\kappa^2 -\xi^2)}\lbrace e^{-\kappa
z}(e^{a(\kappa^2 -\xi^2 )t}-1)+ \nonumber \\ &+&(h+\kappa
)\left[\frac{e^{-\xi z}}{2(h+\xi )}{\rm
erfc}\left(\frac{z}{2\sqrt{at}}-\xi\sqrt{at}\right)\right.
\nonumber \\ &-&\frac{e^{\xi z}}{2(\xi -h)}{\rm
erfc}\left(\frac{z}{2\sqrt{at}}+\xi\sqrt{at}\right)- \nonumber \\
&-&\frac{1}{2}e^{a(\kappa^2 -\xi^2 )t}\left(\frac{e^{-\kappa
z}}{\kappa +h}{\rm
erfc}\left(\frac{z}{2\sqrt{at}}-\kappa\sqrt{at}\right)\right.
\nonumber \\ &-&\frac{e^{\kappa z}}{\kappa -h}{\rm
erfc}\left.\left(\frac{z}{2\sqrt{at}}+\kappa\sqrt{at}\right)\right)
+ \nonumber \\ &+&\frac{h(\kappa^2 -\xi^2 )}{(\xi^2 -h^2
)(\kappa^2 -h^2 )}e^{zh+a(h^2 -\xi^2 )t}\times \nonumber \\
&\times&\left.\left.{\rm
erfc}\left(\frac{z}{2\sqrt{at}}+h\sqrt{at} \right)\right]\right\}
.
\end{eqnarray}

The solution \eqref{TF} of this problem reveal the maximum of the
axis component of thermal-elastic stresses $\sigma_{zz}(0,z,t)$
\cite{12}. This maximum is formed at the depth of the metal to
the end of laser pulse. The value of the maximum, and the value
$z_0$ depends on the parameters of laser irradiation, and the
physical properties of irradiated metal. The value of the maximum
reach the extremum and it position approach to the surface when
the exposure time is decreased \cite{12}. On the surface $z=0$
axis component $\sigma_{zz}(0,0,t) = 0.$ It appears because
replacement of the surface in $z-$up-direction results in a
relaxation of axis stresses. In-plane replacements cause at the
irradiated surface $z=0$ radial-symmetric stresses
$\sigma_{rr},\sigma_{\phi\phi}$ \cite9, which are maximal at the
surface \cite{13} and depend on the depth $z$ as \cite9:
\[\sigma_{rr}=-\frac{E}{1-\nu}\alpha\nabla T(r,z,t),\]
where $\alpha$ is the thermal expansion coefficient; $E$ is the
Young's module; $\nu$ is the Poisson's coefficient.

As it was shown in \cite{12}, stresses, which appeared in
laser-irradiated metal, cause a deformation of a lattice. A
process of the creation of the dislocations will begin in the
volume, which deformation value is more than critical one,
exactly, the yield stress. These areas of deformed bulk of a metal
are determined by the temperature inhomogeneity Eq.\eqref{TF}
\cite5. The processes of the creation of the dislocations are
developed due to an action of the different nature sources,
usually in combination.

\begin{figure}
\hspace{1cm} \psfig{figure=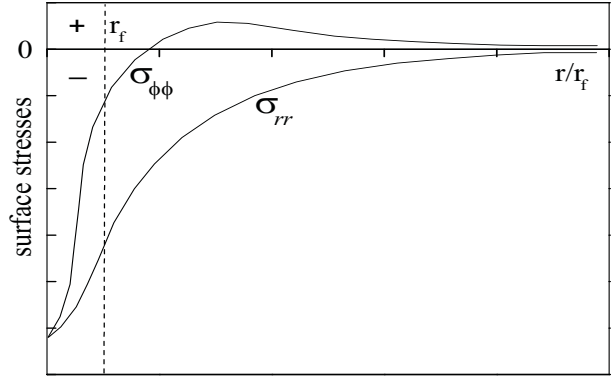,width=8cm,height=5cm}
\caption{Stresses on the surface of a metal, irradiated by a
pulsing laser beam with the spot radius $r_f$} \label{F1}
\end{figure}

Stretched area will appear near the center $(r=0)$ of the laser
spot on the surface $(z=0)$ of the irradiated metal, and the
compressed area will form at some distance from the center of
laser spot. Calculated stresses \cite{12}, which correspond this
influence, are shown on the Fig.\ref{F1}.

New generated, and existent in a bulk of the metal, dislocations
are influenced by the inhomogeneous thermal-elastic stresses. As
a result, effective forces, which appears due to the action of
the inhomogeneous thermal-elastic stresses, cause the
displacements of the part of dislocation or whole piled up
dislocation group, which, in turn cause the plastic deformation of
the near-to-surface areas of the laser-influenced metal. The
transmission electronic microscopy data show \cite{7}, that the
formation of the cellular structure is the result of the creation,
accumulation, redistribution, of the dislocations, on which the
number of the kinks is considerably increased. This formation
depends on the number of the laser shots. Observed experimentally
increasing of the dislocation loops is connected with the
generation of the point defects under described influences.

A motion of the dislocations and its interactions also cause the
generation of point defects \cite4. A type of generated defect,
namely, a vacancy or an interstitial atom, depends on as the
dislocation interaction mechanism \cite{14}, \cite{15}, as the
deformation velocity \cite{16}. Generated vacancies and
interstitial atoms are considerably non-equilibrium defects.
These defects tend to annihilate and disappear in different sorts
of the sinks, such, for example, as the dislocations, a surface,
grain boundaries \cite{17}.

\section{Calculations and estimations}

\begin{figure}
\hspace{1cm} \psfig{figure=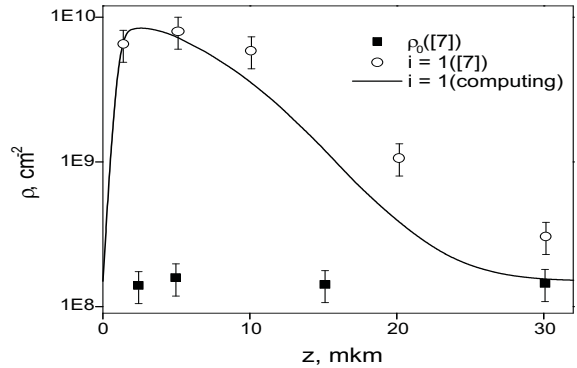,width=8cm,height=5cm}
\caption{Dislocation density in laser irradiated $Armco-Fe.$
Experimental points are obtained at Q-switched laser pulse
duration $\tau =5\times 10^{-8} s.$} \label{F2}
\end{figure}

As it was admitted above, appeared in pulse laser influenced
near-to-surface layers of a metal thermal gradient cause the
thermal stresses. A relaxation of these stresses is realized by
the generation of moving dislocations during the time when the
value of these stresses is more than value of a yield stress
\cite5. Therefore one has taken into account an increasing of the
number of dislocations during as the thermal stresses relaxation
time, as a many-fold action. The analysis in \cite{12} shows,
that the deformation due to the pulse laser irradiation of a
metal is local deformation. A half-width of deformation curve is
greater than a half-width of the laser beam in the case of large
values of $\tau ,$ and difference between these half-widths tends
to zero when $\tau$ is decreased. It is clear, that the defect
distribution will correlate with these tendencies. As it follows
from the results of the works \cite5, \cite9, $z-$axis component
of the density of the dislocations in a metal after laser
irradiation is:
\begin{equation}\label{2}
\rho_{i+1} =\rho_i-\frac{2\alpha}{(1-\nu )b}\int\limits^n_1
\nabla T(0,z,\lambda t)d\lambda ,
\end{equation}
where $i$ is the number of laser shots ($i=0,1,...$); $b$ is the
value of B\"urger's vector of a dislocation; $\rho_0$ is the
initial density of dislocations; by the factor $n$ we take into
account the processes, which delay a creation of dislocations by
thermal stresses.  Calculation of the dislocation density
Eq.\eqref{2} was carried out for the case of $Armco-Fe$ influenced
by Q-switched laser pulse with $q_o=10^7 W/cm^2 ; \tau =50ns;
r{_f}=0.2cm.$ Results of these calculations at $h=0$ are shown on
the Fig.\ref{F2} (continuous line). Experimental data are shown on
the Fig.\ref{F2} also. Experimental points were obtained in \cite7
by a measuring of the density of dislocations by the method of
electron microscopy in $Armco-Fe$ sample, which was laser
irradiated. Calculations were made in supposition of gathering of
the dislocations during relaxation processes. Best correlation of
the calculation and experimental data was achieved at the
relaxation time of order $1.6\mu s.$ Therefore we took this time
as a time of dissipation of the energy, which was pumped by a
laser irradiation into the metal. During this time a generation
and motion of the dislocations are ended.
Small discrepancy between exact solution Eq.\eqref{TF} and real
experimental data one can see in a decreasing part of the curve
Fig.\ref{F2}. This discrepancy occurs due to the volume stresses
$\sigma_{rr}$ and $\sigma_{\phi\phi},$ values of which may be
large enough to create additional notable append to the
dislocation density. We have taken into account also a result of
non-linear depth dependence of a dislocation creation process,
which is facilitated in the less-defective bulk metal region.
Interstitial atoms and vacancies are created during a plastic
deformation of the metal. The difference of its creation energies
in a deformed crystal is not determining factor for concentration
ratio of these point defects \cite{18}. You have taken into
account a dynamics of a dislocation interaction, i.e. it velocity
$V$, stresses $\sigma$ parameters (deformation value
$\varepsilon$, velocity $\dot{\varepsilon}.$) Moving skew
dislocations create interstitial atoms. The velocities of these
dislocations are greater than critical velocity when the stresses
exceed its critical value \cite{19}. It appears on the early
stages of crystal deforming, when the density of the dislocations
is low, and number of the steps, which create interstitial atoms,
exceeds not the number of the steps, which create vacancies
\cite{16}. Point defects in crystals are less movable than the
dislocations \cite{19}. Transfer of the dislocations under
influence of the internal or external forces may results in mass
transfer. Authors of the work \cite{20} supposed, that this mass
transfer is a result of interaction of moving dislocations and
interstitial atoms. Temperature dependence of a preference
parameter for dislocations was calculated in the work \cite{21}.
It was shown, that in the temperature interval $400-900 K$
interstitial atoms appear in the dislocation core, and vacancies
are in its atmosphere. It is clear, that in the case of the edge
dislocation, the space under the dislocation line is preferable
for interstitial atom position. When this dislocation will move in
the stresses field along it glide line, the interstitial atom will
move with dislocation in the place of energy minimum. This atom
transfer will take place until the realization of another, more
energetically preferable situation will realized.

First, atom annihilate on the sink. This case is trivial and we
will not consider it here.

Second, in the case, when the value of the stresses exceeds the
value of Pierl's stress, the non-activated motion of the
dislocations is possible. The velocities of this motion may be
near the sound velocity \cite{22}. At the some critical velocity
value $V_c$ dislocation may loose the atom. Let us estimate this
velocity. We suppose, that a complex {\it "dislocation +
interstitial atom"} will be exist until it kinetic energy is less
than the link energy of the interstitial atom with the
dislocation. We can write:
\begin{equation}\label{3}
\frac{1}{2}MV_c^2\geq W,
\end{equation}
where M = m* + m, m is the atom mass; m* is the dislocation
effective mass, which is determined as [19]:
\begin{equation}\label{4}
 m*=m\frac{2d}{l},
\end{equation}
where $d$ is the distance between glide planes; $l$ is the
half-width of a dislocation: $$l=\frac{d(3-2\nu )}{4(1-\nu )}.$$
Hence we get:
\begin{equation}\label{5}
V_c =\sqrt{\frac{2W(3-2\nu )}{m(11-10\nu )}}.
\end{equation}
In the case of the $Armco-Fe$: $W = 0.2\div 0.5 eV$ \cite{23};
$\nu=0.28; m = 1.66\times 10^{-24}M_{Fe}=9.27\times 10^{-23}g,$
and
$$V_c=(450\div 716)\times 10^2 cm/s$$

Now it is easy to see, that in the case of free motion with the
velocity $V_c$ during the time $t_M=1.6\times 10^{-6}s$
(relaxation time) complex {\it "dislocation + interstitial atom"}
will be on the distance of $720\div 1146 \mu m.$ Experimentally
measured depth of the atom penetration from initial layer of
$1\mu m$ is in order of $10\mu m$ \cite1, \cite2 (at Q-switched
laser pulse duration $\tau =3\times 10^{-8} s$), which is
considerably less than the above mentioned distance. This
difference is caused by an interaction of the dislocations with
different defects during a motion through the real crystal. A
dissipation of the complex energy results in a decreasing of the
velocity of the complex. The defects, which cause energy
dissipation in real crystals, are other dislocations, point
defects, disperse constituents of other phase, etc. \cite{19}. To
estimate these losses let us consider the motion of the complex
{\it "dislocation + interstitial atom"}, which mass is $M,$ under
the influence of the stresses forces $F = f(\sigma ),$ in the
media with the damping coefficient $\gamma .$

During a laser pulse ($t\leq\tau$) motion equation
\begin{equation}
\label{6} d_t V_1=\frac{F}{M}-\frac{\gamma}{M}V_1
\end{equation}
solution is:
\begin{equation}
\label{7} V_1=\frac{F}{\gamma}(1-\exp\{-\frac{\gamma t}{M}\}),
\end{equation}
which presents maximum velocity
\begin{equation}
\label{8} V_{max} =\frac{F}{\gamma}(1-\exp\{-\frac{\gamma
\tau}{M}\}).
\end{equation}

When the stresses are not influence the motion under consideration,
in other words, when the laser pulse is ended, a complex {\it
"dislocation + interstitial atom"} reaches the maximal value of the
velocity, which is described by the equation
\begin{equation}
\label{9} d_t V_{2}=-\frac{\gamma}{M}V_2 ,
\end{equation}
which solution is
\begin{equation}
\label{10} V_2=V_{max}\exp\{\frac{\gamma}{M}(\tau -t)\}.
\end{equation}

Averaging of the given solutions over the relaxation time $t_M$
$$<V> =\frac{1}{t_M}\int V(t)dt$$
gives
\begin{equation}
\label{12} <V>=\frac{F}{\gamma t_M}\{\tau
+\frac{M}{\gamma}[1-\exp\{\frac{\gamma
\tau}{M}\}]\exp\{-\frac{\gamma t_M}{M}\}\}
\end{equation}

In the case of large values of $\gamma$
$$<V> \approx\frac{F\tau}{\gamma t_M},$$
and a ratio
\begin{equation}
\label{13} \frac{<V>}{V^{(0)}_{max}}\approx\frac{\tau}{t_M}\approx
0.02.
\end{equation}
Here $$V^{(0)}_{max}\equiv\frac{F}{\gamma}.$$

Taking into account given ratio, and putting $V^{(0)}_{max}=V_c ,$
we can get for a real depth of surface atoms mass transfer the
value $$z = <V>t_M \approx V_c\tau\sim 14\div 22\mu m.$$

{\it "Dislocation + interstitial"} complex will accelerate to the
velocity of the order $V_c$ at the initial period of a laser
influence, i.e. at the times $\tau /2$ (if the Q-switched laser
pulse form is near to triangle). It gives complex decomposition
depth $\sim 7\div 11 \mu m,$ which is in a good agreement with the
experimental data. Better correlation is achieved by taking into
account the possibility of the annihilation of the dislocations.
The dislocation annihilation takes place when the opposite sign
dislocations collide in the process of the motion in parallel
glide planes \cite{15}. Simplest annihilation realized for the
dislocations with a free core. In the case of the {\it
"dislocation+interstitial atom"} complex annihilation is
low-possible because interstitial atoms in the core of dislocation
enable this effect. For an edge dislocation another velocity
decreasing factors may be the delay of the kinks on the steps of a
dislocation, and on the vacancies, which are situated on the edge
of the extra plane.

We have admit, that the tagged atoms (radioactive isotopes of $Fe$) are
concentrated in smooth near-to surface layer of experimentally investigated
samples. Therefore the ratio of penetration depth of tagged atoms to whole
depth of high dislocation concentration is proportional to initial ratio
of value of tagged atoms layer and value of influenced thickness of a sample.

We can estimate the depth of vacancy atmosphere loosing by a
dislocation, using the Eqs. \eqref{5}, and \eqref{13}. These
events will be realized on the depth $\sim 3\mu m$ if we put the
{\it "vacancy+edge dislocation"} binding energy value equal to
$0.04eV$ \cite{23}. It is clear, that binding energy {\it
"dislocation + interstitial atom"} exceeds the binding energy {\it
"dislocation + vacancy"}, therefore during the motion of complexes
with the velocities, values of which are more than the critical
value, the dislocations loose first of all the vacancy atmosphere
and then the interstitial atoms, which are bounded to the core of
the dislocations. This difference in the behavior of the vacancies
and interstitials is thermal activated and realizes in the
considered temperature interval. Calculated in \cite5 temperatures
in the metal, which was influenced by pulsing laser irradiation,
are in good agreement with this idea. Calculations in \cite5 and
experimental data in \cite{24} show that temperature influence of
the laser beam during nano-second reaches the depths in order of
$\mu m.$ As a result the concentration of the vacancies in the
near-to-surface layers are much than the concentration of the
interstitials and vise versa in deepest layers \cite4 .

\section{Discussion}

Non monotonous dependencies of the distribution of tagged atoms
under surface of pulse laser irradiated metal are in good agreement
with the calculations, which were carried out on the base of a
concept of the presented above mechanism of capturing and transport
of the point defects by the dislocations, which are generated and
moving by the thermal fields, which appeared in surface layer of
irradiated metal.

The kinetic of a gathering of the tagged atoms
$^{55}Fe$+$^{59}Fe,$ which were penetrated from the surface to
bulk of a metal, which was influenced by many-fold pulse laser
irradiation ($\tau\sim 10^{-8}s, W\approx 0.02 J,$) was
investigated in $Armco-Fe$ \cite{25}, \cite{25a}. A maximum of the
concentration of radioactive atoms $^{55}Fe+^{59}Fe$ began to be
formed after the first three cycles of an irradiation. It
correlates with $X$-ray and electron microscopy data, which shows
maximal increase of defects in a metal after the first cycles of
pulse laser influence. A displacement of the defect concentration
maximum to the depth of the irradiated sample was observed
simultaneously with the forming of the maximum. These processes
saturation are observed at the number of laser shots $\sim 10.$
At this time an increasing of the dislocation density in
$Armco-Fe$ was ended, and new dislocations were gathered into
cellular structure, as it was shown by the electron microscopy
experimental data \cite7 .

Effective diffusion coefficients $D^*_i$ were obtained from the
analysis of the tagged atoms concentration experimental data,
which were measured in the influenced by pulsing many-fold laser
irradiation areas of $Armco-Fe.$ The value of $D^*_i$ is monotony
decreased with the number of laser shot $i$ \cite{24}. This effect
is caused by an increasing of the number of the defects, which are
created by laser influence. It results in the effective slowing
of the motion of the complexes {\it dislocation + interstitial
atom}. A value of $D^*_i$ is decreased not so fast after $5\div
8$ shots. This slowing is connected with a beginning of the
process of effective redistribution of the radioactive isotope
\cite7 .

We can describe the experimental data of the works \cite1, \cite2
from the positions of the presented above {\it dislocation +
interstitial} mechanism taking into account the data of the work
\cite{26}. Compression wave appears in the metal during the time
of influence of pulsing laser irradiation. Stresses wave slope is
increased during this laser influence on a metal. A number of
defects of crystalline lattice is increased due to appearance of
the inhomogeneous stresses as a result of laser influence. New
defects may be the sinks for migrating from the surface
interstitials. Tagged atoms concentration is maximal on the depth
$z_0\geq\sqrt{a\tau},$ which coincides with the depth of maximum
of thermal stresses at the moment of the end of laser pulse. The
number of sinks is decreased with the depth because the distance
to the source of the stresses increased, and amplitude of the
stresses is decreased. Each next laser pulse pumps radioactive
isotopes from the near-to-surface layers to the bulk of the
metal, and increases the number of sinks for interstitials. It is
reflected on the slopes decreasing of the exponential parts of
tagged atom concentration curves as a function of $i$ \cite2.
Hence a decrease of the value of $D^*$ is slowing.

Presented above phenomenological description of the
laser-stimulated {\it dislocation + interstitial} mass transfer
process in $Armco-Fe$ is confirmed by the experimental data of
the investigation of self atom transport in $Ni$ \cite{27}. It
appeared, that increasing of the vacancy number slowing mass
transfer. This phenomenon is not possible to explain in the frame
of the vacancy mechanism because an increasing of the number of
the vacancies facilitates a diffusion of the atoms in a
crystalline lattice. Observed phenomenon is explained easy by the
{\it dislocation + interstitial} mechanism of a pulse influenced
mass transfer. At the beginning of a mass transfer migrating
interstitials are high movable and interact not practically with
the vacancies, which are traps for its. Loosing of a kinetic
energy by the interstitials makes its more suitable for vacancy
trapping, by the other words a recombination probability
increase. This recombination reflects a mass transfer last
stadium, which is characterized by exponential dropping parts of
the concentration curves.

Presented above estimations and analysis show the nature of
pulsing laser stimulated mass transfer in metals. This mass
transfer is carried out by the near-to surface layers
interstitials, which interact with the moving dislocations during
the laser influence and relaxation of stressed state of a
crystalline lattice. Taking into account that the stressed state
relaxation processes are common for the all pulsing influences
\cite{28}, which not destroy the structure of an influenced
metal, we can state an universality of the presented mass transfer
mechanism.

\end{document}